# Three-Dimensional Spatial Cognition: Bees and Bats

Robert Worden

Theoretical Neurobiology Group, University College London, London, United Kingdom

[rpworden@me.com](mailto:rpworden@me.com)

Draft 1.8; May 2024

**Abstract:**

Every animal needs to understand the local space around it, to control its movements. A Bayesian analysis of cognition defines the best possible 3-D model of space which can be built from sense data of limited resolution. Computing this optimal model is too costly for animals to do in real time, but it can be done on computers. The paper describes a program which computes the best possible Bayesian model of 3-D space from vision (in bees) or echo-location (in bats), at Marr's [1982] Level 2. The model exploits the strong Bayesian prior probability that most other things do not move, as the animal moves. 3-D locations of things are computed from successive sightings or echoes, computing locations from the animal's motion. The program can be downloaded and run, from www.bayeslanguage.org/bb/BB.zip .

The full Bayesian computation is not tractable for animals. They could use a tracking approximation, which only requires storage of the latest position estimate for each object they track. The program computes the tracking approximation. If memory for positions does not add errors, tracking gives a 3-D model close to the best possible model. Both models require short-term memory for 3-D locations. The program computes how memory errors affect tracking. If spatial memory is stored as neural firing rates, expected error levels greatly degrade the quality of the tracking model.

An important use of the 3-D internal spatial model is to detect other moving things, as the animal itself moves. Sometimes motion cannot be detected directly from the visual field, due to the animal's own motion; a 3-D spatial model of locations is required. Motion detection is modelled in the program. Expected levels of neural memory errors sharply degrade motion detection.

Neural implementations of the spatial model face a major challenge, that neural short-term spatial memory is expected to be too imprecise and too slow. Something else is required. A 3-D model of space could be stored in a wave excitation, as a Fourier transform of real space. This could give high memory capacity and precision, with very low spatial distortion, fast response, and simpler computation. Evidence is summarized from related papers that a wave excitation holds spatial memory in the mammalian thalamus, and in the central body of the insect brain.

# 1. Introduction

Animal brains make internal models of their surroundings. In the Bayesian analysis of cognition [Worden 2024a], it can be shown that in complex domains, animals need to use internal Bayesian models to give the greatest possible fitness. All animals need an internal model of the 3-D space around them, to control their movements at any moment of the day. This gives large and sustained selection pressure to make the 3-D model of local space as precise as their sense data permits.

There is a fittest possible internal spatial model that can be built from sense data of limited precision. This best model can be computed by a Bayesian computation, which is too costly for animal brains do in real time, but which can be done in digital computers.

Three reasons imply that for many animal species, their internal model of space is very nearly the best possible spatial model, given their sense data:

1. Informal observations of many species suggest this: for instance, even a tiny insect can land on an irregular surface with speed and precision.
2. Spatial cognition has a huge impact on lifetime fitness. If some species' internal model of space was not near to the best possible, any other species with a better internal model would have a huge fitness advantage over it, and would drive it to extinction.
3. Animals would not invest costly resources in gathering precise sense data, if they could not make the very best use of it with their brains.

This leads to a **Spatial Modelling Hypothesis**:

**Animals' internal models of local space are close to the best possible model that can be inferred from their sense data.**

This paper examines the hypothesis:

- Section 2 summarises a related paper [Worden 2024a], which shows that for a given set of sense data, there is a best possible internal model of 3-D space, which can be computed using Bayes' Theorem. It describes the selection pressures which have acted consistently on animals since the Cambrian era, to make their internal model of local three-dimensional space as precise as it can be.
- Section 3 describes how internal 3-D models of space can be built from sense data. The main technique used in this paper is to compute object locations from the animal's own motion (Structure from Morion, SFM [Murray er al 2003]), using the strong Bayesian prior probability that other objects do not move as the animal moves.
- Section 4 describes the program which uses SFM to compute a Bayesian optimum spatial model, at Marr's [1982] Level 2. The program uses discrete time-steps. At each step the animal uses sense data from the last few steps to compute the most likely positions of objects. Program results are shown for bees (vision) and bats (echo-location).
- Section 5 describes a tracking model, in which the animal compares its current sense data with its previous best estimated position, to track the positions of objects. If the estimates are stored with high precision, tracking gives almost the same 3-D model as the optimal Bayesian model.
- Section 6 explores the effect of limited-precision memory of position estimates, on the precision of the tracking. Tracking precision degrades as random short term memory storage errors are increased. Tracking errors grow to be larger than the memory errors.
- Section 7 analyses the problem of detecting moving objects while the animal moves, which drives large competitive selection pressures. The tracking models are used for motion detection. The effectiveness of movement detection is sharply reduced by errors in spatial memory.
- Section 8 describes how object tracking could be used in an aggregator computation of 3-D space, combining stereoscopy, multi-sensory integration and object recognition.
- Section 9 explores how the 3-D model could be neurally computed in animal brains. There is a major difficulty, that neural storage of spatial information cannot give the required precision in sub-second timescales. This challenge could be explored in neural processing frameworks such as the Free Energy Principle.
- Section 10 introduces an alternative to neural storage of the spatial model, in which the location vectors of objects (and their uncertainty tensors) are stored in a wave excitation which couples to neurons. A wave has advantages over neural storage – of high memory capacity, high spatial resolution, low spatial distortion, and simple computation.
- Section 11 discusses how the wave might be realized in animal brains. The wave excitation may be held in the mammalian thalamus, and in the

- central body of the insect brain. Evidence for a wave from related papers is summarized.
- Section 12 discusses the implications of the wave hypothesis for theories of consciousness. Consciousness includes a largely veridical model of local 3-D space. This is the most important empirical data about consciousness, and is the most stringent test of any theory of consciousness. If consciousness arises from a wave, it could fit this data in a simple way.
- Section 13 responds to papers by Hoffman et al. on the Interface Theory of Perception. These papers propose that animals' internal models of 3-D space, derived from vision, are not veridical models. A recent paper 'Fitness Beats Truth' (FBT), proves that animals which only act to maximise their fitness always out-compete Bayesian animals (which build veridical internal models), and drive them to extinction. This section shows that fitness beats truth only when animals have very few choices of action; and that for the many choices of physical motion supported by vision, fitness needs truth.
- Section 14 collects the main results of the paper, and discusses directions for future research.

The figures in this paper are drawn by the interactive demonstration program mentioned above. The program can be downloaded and run from www.bayeslanguage.org/bb/BB.zip to explore how the performance of 3-D cognitive spatial models depends on parameters such as visual acuity and memory noise.

A key result from this modelling work is the proposal that there may be a wave excitation in the brain, holding 3-D spatial information. There is other evidence for this hypothesis [Worden 2020a, 2024b], which is now strong enough that it is worth exploring.

Researchers might hesitate to join a search for a wave in the brain, because the idea is unconventional. However, conventional neuroscience has continued for many years; the main research topics are well-trodden paths, explored by large teams. For a young researcher, looking for a wave in the brain is an unexplored green-fields area, where new ideas can be proposed and where discoveries are to be made. If the wave hypothesis were confirmed, it would be no less than an earthquake in neuroscience. It is worth the attempt.

## 2. Optimal Bayesian Cognitive Models

It is known [Cox 1961] that Bayesian models of decision making are optimal in some settings. It can be shown [Worden 1995, 2024a] that Bayesian cognition, using prior probabilities that match the actual probabilities in an animal's habitat, gives the greatest possible fitness. Through natural selection, animal cognition converges towards this Bayesian optimum.

[Worden 2024a] shows that for maximum fitness, an animal needs to choose its actions **as if** it had made the Bayesian computation. For simple choices of action, it may not be necessary to make the Bayesian calculation; some simpler short-cut computation may give the same choices of action, and so give the same fitness. Spatial cognition is not one of these simple cases, as it underpins every choice of actions that an animal makes. There are no short cuts, and an animal needs to build the best internal model of space that it can, using its sense data.

The Bayesian prior probabilities in an animal's habitat, which define the optimal internal 3-D model of local space, are of two kinds:

a) The probability that objects in local space obey the constraints[1] of 3-D Euclidean geometry, kinematics and physics
b) The probabilities that different types of object, with different visual or sensory signatures, move in certain ways or have certain spatial forms.

Probabilities of type (a) have been true for all evolutionary time – for more than half a billion years, since the Cambrian era, when capable spatial cognition became necessary for many species. Probabilities of type (b) have fluctuated, over periods as short as a few years, for instance whenever a species moves to a new habitat. The selection pressures on brains from the constraints of type (a) have been constant over a huge time period; whereas the pressures of type (b) have been transient and fluctuating.

Animal cognition converges towards the Bayesian optimum, but at a limited speed, through random genetic changes [Worden 1995, 2023]. So we expect cognition to have responded to pressures of type (a) quite precisely, but much less precisely to type (b) pressures.

For over 500 million years, many animals have had complex sense data (such as precise vision) and capable limbs. At every waking moment, they need to control their limbs skillfully. To do this, they need to understand the local 3-D space around them, over a range of scales – from proximate space where their limbs are, to more distant space, containing things they need to reach or use or evade.

---

[1] Geometric constraints are not usually thought of in terms of probabilities, but they can be seen that way – as Dirac delta functions in a 3-D probability density defined over space.

The costs of any mistake in doing these things, arising from an inaccurate internal 3-D model of the world, can be large – amounting to a high risk of death at any moment of the day. Because the internal 3-D model of space is tested every second of the day, the lifetime selection pressure to make it as accurate as possible is very large – larger than the selection pressure on making less frequent choices[2]. This huge selection pressure has continued for more than 500 million years. There is no other aspect of brain function that has been under any comparable selection pressure, compared to that which has shaped spatial cognition.

This may mean that more than 95% of the consistent and persistent selection pressure on the brain has been the pressure to refine 3-D spatial cognition. Other selection pressures have been either weaker (pertaining to less frequent events in an animal's life), or fluctuating over evolutionary time, or both. Spatial cognition is the pre-eminent thing that brains need to do well.

That is the rationale for the spatial modelling hypothesis of this paper - that the internal 3-D model of local space comes very close to the Bayesian optimal model. If it did not, then any species which evolved to have a better spatial model would have a huge selective advantage over other species, and would drive them to extinction.

The problem of spatial cognition is so important that it is worth devoting a lot of brainpower – a lot of design complexity, and a lot of energy – to doing it as well as possible. That is why, amongst all forms of cognition, spatial cognition is primary – the first call on an animal's brain, which all other types of cognition depend on. Whatever an animal needs to do, it needs to understand space in order to move to do it. If cognition is like an atom, spatial cognition is the nucleus of the atom.

A possible reason for the spatial modelling hypothesis to be wrong – for animals' internal 3-D spatial models to be far from the optimum - would be that a close approach to the optimum is truly impossible; if there is some impenetrable barrier in the design of brains, which the evolution of brain designs can never get past. As will be discussed later in the paper, there may be such a barrier, for a purely neural brain; but there is a way round it, which evolution may have found.

There is empirical evidence that evolution has found a way; that in many animals such as bees and bats, the brain's internal model of space is very good indeed, and is very nearly as good as their sense data allows. This hypothesis is subject to further and more precise empirical tests.

## 3. Building Internal Spatial Models

Suppose an animal is moving near a small stationary object. Before the animal first sees the object, it knows nothing about the object's location. In Bayesian terms, its ignorance is represented as a uniform prior probability density P(**r**) for the location **r** of the object, in some region of space containing the animal and the object. **r** is a 3-vector position.

When it first sees the object, the object's location is constrained to be approximately along its line of sight. If the animal's vision (denoted by v) has limited resolution, the conditional probability density P(v|**r**) is a narrow Gaussian cone in 3-D space around the line of sight. By Bayes' theorem, the posterior probability density in **r** is given by

$$P(\mathbf{r}|v) = \text{const}*P(\mathbf{r})*P(v|\mathbf{r})$$

As the animal moves, the object appears to move in the animal's visual field. In the modelling of this paper, the animal's motion is divided into a series of short time steps. For each time step, by Bayes' theorem, the posterior probability density of the object's location after the step is the probability density from the previous time step, multiplied by the narrow Gaussian cone from the latest sense data. Over a series of steps, a number of Gaussian cones in 3-D space - one for each time step - are multiplied together. Multiplying several narrow Gaussian cones allows the animal to take a series of cross-bearings on the object, and hence to know its three-dimensional position relative to the animal.

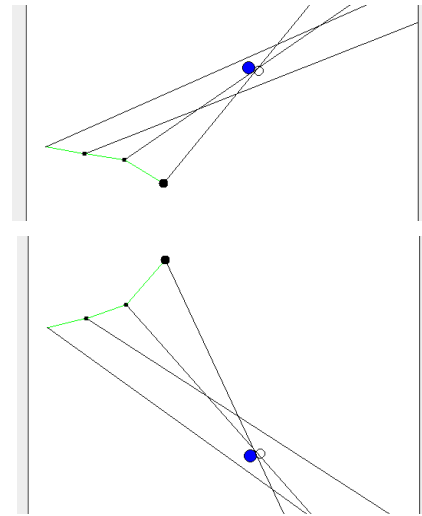

*Figure 1: Estimating the three-dimensional position of an object, for a bee. Successive lines of sight to one object are shown. The figure shows two views of the same three-dimensional scene, from different angles. Errors in the lines of sight are exaggerated for clarity.*

---

[2] The strength of the selection pressure to choose some action correctly is proportional to the number of times that choice must be made in a lifetime, times the fitness cost of making it incorrectly

Figure 1 is taken from the demonstration program described in this paper. It shows two views from different angles of the same three-dimensional scene, as displayed by the program. The trajectory of the bee is shown in green, and the true position of the object is the blue circle. Lines of sight from the bee to the object are black lines; because of the limited resolution of bee's vision, the lines have angular errors (exaggerated for illustration) and do not pass exactly through the object.

The Bayesian maximum likelihood position of the object is shown by the white circle. This is found by multiplying the Gaussian cones for the lines of sight, and finding the maximum of the product. This can be done by summing negative log likelihoods, and finding the minimum. In the program, this is a computation of 3*3 matrices, iterating rapidly to the minimum point. It is a Newton-Raphson iteration in three dimensions.

For a bat using echo-location instead of vision, the sense data at any one time step gives two constraints on the location of each object:

a) The time delay of the echo measures the distance from the bat to the object – and so it constrains each object to the surface of a sphere centred on the bat
b) For any stationary object, the doppler shift of the echo (from the bat's motion) varies as the cosine of the angle between the bat's direction of motion and the object's position relative to the bat, and so the doppler shift measures this angle. This constrains the position on the sphere from (a), to be on a circle in 3-space.

As the bat moves, if its sense data was precise, the successive circles would all intersect at the position of the object. The sense data are not precise, but have errors (modelled as Gaussian 'doughnut' distributions around the circles); so the circles do not intersect, as is shown in figure 2. A maximum likelihood position can be estimated, as for the bee.

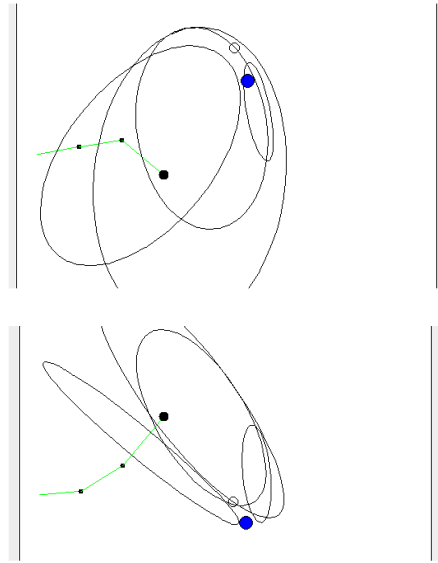

*Figure 2: Estimating the three-dimensional position of an object, for a bat. In each time step, echo-location constrains an object's position to be on a circle in space. Circles from several times steps are shown. All circles pass approximately through the object's location, but because of random errors they do not intersect precisely.*

The path of the bat is shown in green, and the object is blue. The circles of constraint are shown in black, and the most likely inferred object position is the small open circle.

When there are many objects in vision or within echo-location, the same computation can be done for each object; the amount of computation (and the memory required) increases only linearly with the number of objects.

The calculation depends on a strong Bayesian prior probability that objects do not move, as the animal moves. It is a computation of Structure From Motion (SFM), treating the objects as the structure of a rigid body. For a fast-moving animal, this assumption is approximately correct for many types of object.

To construct the optimal (most precise) 3-D model of space involves retaining and using previous sense data for variable time periods, depending on the type of object and how likely it is to move. This full Bayesian modelling (which the program does) requires retention and use of precise sense data for several previous time steps, so is demanding in computation and memory. We do not expect animal brains to do this; we expect them to use some simpler approximation, if it is feasible and if it gives similar results to the true optimum computation.

There is a good approximation to full Bayesian spatial cognition, called the **tracking approximation**, which takes account of the geometric constraints of type (a), but takes more limited account of type (b) factors. In this approximation, computation of the 3-D model at each step does not use sense data from previous steps; it only uses sense data from the current step, combined with the result of computing the locations at the previous step. For each object, rather than estimating its location *ab initio* at each step, combining sense data from several previous steps, the tracking approximation just combines the previous estimate with current sense data. This reduces the requirement for short term memory and computation.  The program computes the tracking approximation.

What about uncertainty in the animal's own location, orientation and speed? How should these affect the computation? By Galilean invariance, the animal's location and motion is only defined relative to local objects. The problem of finding a moving animal's location is mathematically equivalent to the converse problem, in which the animal does not move, but local objects move. If many objects are stationary relative to the ground, they effectively form one large rigid body; the animal is estimating the shape and position of the rigid body relative to itself, using a structure from motion calculation.

[Fields 2023] has characterized simple animal cognition about spacetime as a cycle:

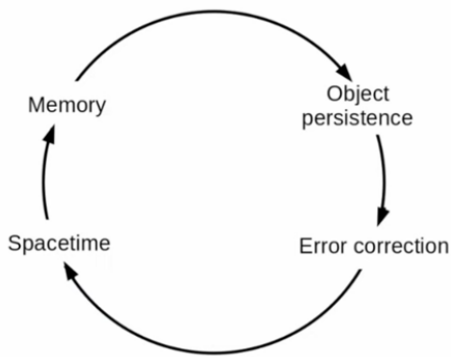

*Figure 3: Cyclic model of cognition, from Fields [2023]*

The spatial models in this paper use this cycle. For instance, in the tracking model:

1. **Object persistence** is needed to recognize that sensory input comes from the same object in successive time steps; that an object seen now is the same object as one in a previously computed 3-D spatial model
2. **Error correction** involves taking an estimate of an object's position from the previous step, and updating it using current sense data, to minimize discrepancies
3. **Spacetime,** as a time-dependent 3-D model of local space, is the result of the computation
4. **Memory** is required to persist the spatial model from one time step to the next. For each object, a 3-vector of its position, and a 3*3 tensor of the uncertainties in its position, is retained.

The program computes locations from motion, which is possible for any animal with vision or echo-location. This could serve as a foundation for multi-sensory integration.

## 4. Computing Bayesian Optimal 3-D Models

The program computes three different models of 3-D space, for bees (vision) or bats (echo-location):

1. **The Optimal Full Bayesian Model**: direct use of Bayes' theorem to combine sense data from the current step and several previous steps, to build the best possible 3-D spatial model from sense data.
2. **A Tracking Model**: where sense data from the current time step is combined with the estimates made in the previous step.
3. **A Noisy Tracking Model**: with variable memory storage errors (noise) – which otherwise, is the same tracking as model (2)

This section describes the full Bayesian model (1), and some features common to all three models.

The animal (a bee or a bat) is modelled as flying in a semi-random manner in a cubical space of side 2.0 – say, 2 meters. Objects are denoted by coloured circles and are placed randomly in the cube. Usually (unless using the program to test motion detection, as discussed in section 7) the objects do not move. The animal estimates object locations in an allocentric frame of reference, in which most objects do not move.

At each step in its track, the animal receives sense data about each object, with Gaussian random errors to simulate the limited precision of its vision or echo-location. The sense data constrains the object's location, maximising the Bayesian likelihood in a Gaussian cone around the line of sight (for vision) or torus/doughnut around a circle (for echo-location).

The animal can recognize persistent objects from one step to the next. The location of each object is computed by multiplying the Bayesian probability densities from sense data in the current step and a number of previous steps. Sense data from earlier steps is downgraded by a factor exponential in the number of steps, to reflect the Bayesian probability that an object might have moved.

This computation is done for each step, separately for all objects that the animal can detect. (the model assumes that by taking an average of all objects, the animal knows its own position and velocity relative to them with high precision)

The computation of each object's likely position and uncertainty, in the approximation that the likelihood function is Gaussian in three dimensions, requires only simple 3-D matrix manipulation and a rapid iteration to the most likely position. It can be done very fast on any PC. The animal's internal 3-D model of space can be displayed, and rotated to show object depths, as in Figure 4:

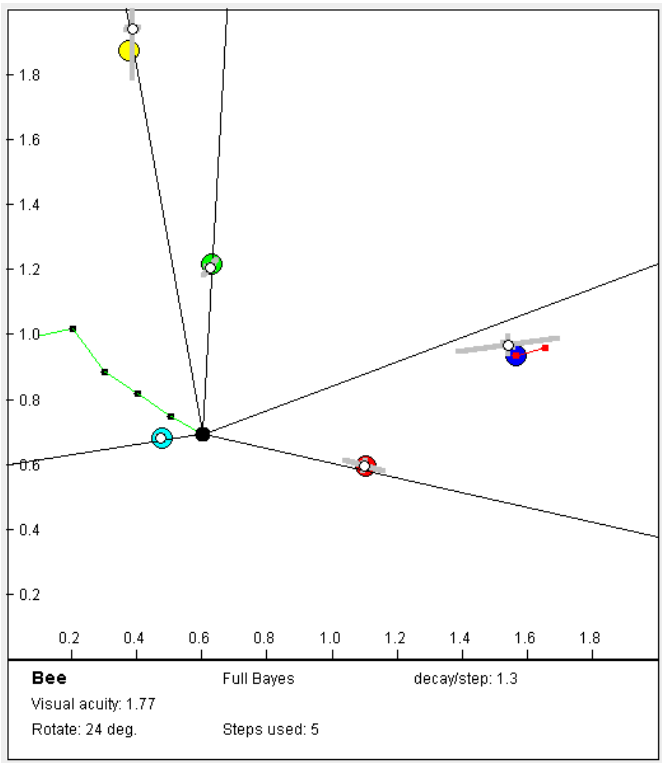

*Figure 4: Program display of a bee's 3-D model of space*

In this figure:

- The steps in the bee's track are the short green lines
- True object positions are shown as coloured circles
- Estimated object positions are shown as small white circles
- The Gaussian likelihood distribution for each estimated object position is shown as a three-dimensional grey cross, with one axis for each eigenvector of the error matrix (each error bar is one standard deviation)
- Lines from the bee to the objects are lines of sight, with Gaussian random errors in direction.
- The area below the window shows some parameters of this run of the model

With this visual acuity (1.77 degrees, which is quite realistic for a bee), the three-dimensional locations of most of the objects are quite well constrained. The largest uncertainty is in the range of any object from the bee, especially if the bee is flying straight towards it.

For the bat, the 3-D diagrams are similar, but with lines of sight replaced by circles (doughnuts) of echo-location.

If the objects are assumed to have small probability of moving, the precision of the location estimates continues to increase over several steps – resulting in higher precision in the 3-D model than in the sense data – because sense data from several steps, with independent errors, has been combined.

This is the first result of the Bayesian model – that by combining sense data from several, a fairly precise 3-D model of object locations can be built. Estimated positions converge towards actual positions, and the size of the error bars reduces.

Typically, the largest uncertainty (for vision) is uncertainty in range of an object from the bee. By clicking the mouse on any object, the likelihood distributions for its range are displayed. An example is shown in Figure 5:

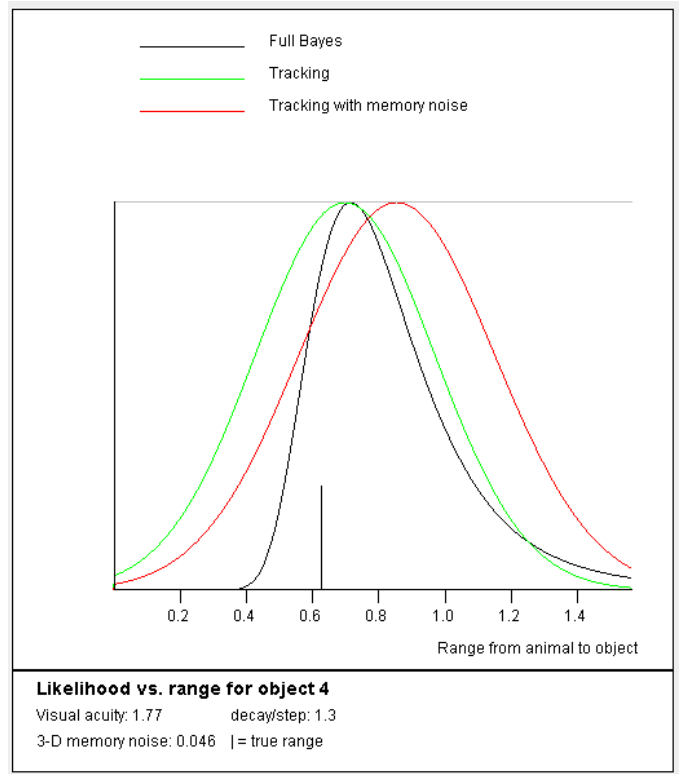

*Figure 5: Likelihood distributions for the range from the bee to one object (un-normalised)*

The black curve shows the likelihood distribution for the full Bayesian calculation. The vertical bar shows the true range of the object. When the range has large uncertainty, the likelihood distribution is asymmetric, and has a longer tail at larger ranges[3].

Parameters of the model can be varied using sliders, to test their effect. Variable parameters include:

- The visual acuity of the bee, or the echo-location precisions of the bat

---

[3] The likelihood distribution of (1/range) is nearly symmetric (as in stereoscopy), which gives an asymmetric likelihood distribution in range.

- The number of steps before (and including) the current step, whose sense data are combined in the Bayesian estimate.
- The decay factor in negative log likelihood per step in previous steps, to reflect the probability that an object may have moved
- The number of objects in the volume
- Parameters of the bee's motion: step length, initial vertical velocity, and random jitter in step direction
- Rotation of the view, about a horizontal axis

In the full Bayesian model, memory for sense data from previous steps, and the computations of maximum likelihood positions, are assumed to be precise (done with Java floating point arithmetic). The sense data in any step has errors, but these errors do not change when the same sense data is reused in later steps (there are no memory storage and retrieval errors).

## 5. Tracking Objects

The program computes a tracking model, in which sense data for any step are not retained in short term memory, but the estimated object positions, and their uncertainty tensors, are retained for one step only.

At each step, the location of each object is tracked, by Bayesian multiplication of its previous location likelihood estimate (in a Gaussian approximation), with the likelihood distribution from the current sense data. In this way, the new maximum likelihood location, and its Gaussian uncertainty tensor, are re-estimated at each step.

Because sense data are not retained and reused over several steps, the tracking model is less demanding in memory and computation than the full Bayesian model, and is a more likely candidate for how animal brains estimate locations from sense data. In the program, as with the full Bayesian model, tracking can be done rapidly by 3*3 matrix manipulation and a fast iteration, separately for each object. The program computes both Bayesian and tracking models at the same time.

The tracking and the full Bayesian calculation are not identical in effect, because the tracking model uses a Gaussian approximation for uncertainties of position estimates. Nevertheless, for typical parameters of the models, the two calculations give very similar estimates and uncertainties. This can be seen by switching the display between the Full Bayesian model and the Tracking model. Switching shows that the estimated object locations, and their error bars, move very little.

This is the main result of the tracking model: that in the absence of memory errors, tracking gives a 3-D model of space which is very similar to the full Bayesian model – that is, similar to the best possible model that can be built from sense data, but requiring fewer computational resources. So tracking is a good candidate for the way animals model objects in space.

The similarity of tracking and Bayesian models can be seen in figure 5, where the maximum likelihood peaks of the full Bayesian range estimate (black curve) and tracking range estimate (green curve) are both close to the true object range. The tracking uncertainty is Gaussian, while the Bayesian uncertainty is asymmetric.

Like the Full Bayesian model, the tracking model assumes that the memory for 3-D object locations has high precision. That assumption is not realistic, if object locations are stored in neural memory; the effect of spatial memory errors is described next.

## 6. The Effect of Memory Errors on Tracking

The program computes a tracking model with memory errors (noise). This is the same as the first tracking model, except that when each object location estimate is stored at each step, random errors with a Gaussian distribution are added. This models the effect of inaccuracies in memory storage and retrieval. The standard deviation of the Gaussian memory error can be varied, and the program models four types of memory noise:

1. **Absolute Range Errors**: an error only in the estimated range of the object, whose size is proportional to the range from the animal to the object
2. **Absolute 3-D errors**: a different random error in all three dimensions, whose size is proportional to the range from the animal.
3. **Relative Range Errors**: an error in the estimated range of the object, proportional to the range from the animal. Errors in nearby objects are correlated, so that the error in the relative range displacement of two objects is proportional to the distance between them.
4. **Relative 3-D errors**: a random error in all three dimensions, proportional to the range from the animal. Errors in nearby objects are correlated, so that the error in the relative displacement of two objects is proportional to the distance between them

In all four types of memory error, objects near the animal have smaller location memory errors.

In the relative error types, the relative locations of nearby objects are stored in a tree-like (hierarchical) memory structure, with nearby objects close in the tree structure, building up displacements by traversing the tree. For any setting of the error size parameter, relative errors are normalized so that the total variance summed over all points is the same as for the absolute error types.

The range error types (1) and (3) correspond to an assumption (possibly unrealistic) that position estimates are stored in the visual cortex, and have small lateral errors, with only a significant range error.

The three-dimensional error types assume that three-dimensional location estimates (which may be stored anywhere in the brain) and are subject to errors in all three dimensions.

The program computes both tracking and noisy tracking (with memory errors). For noisy tracking, it can model any of the four memory error types, to see if they have different effects. The differences it finds between the results for different types are not large.

The effect of memory noise on tracking can be tested by switching the view from tracking to noisy tracking, to see how estimated object positions move. They may move significantly, as is shown in figure 6.

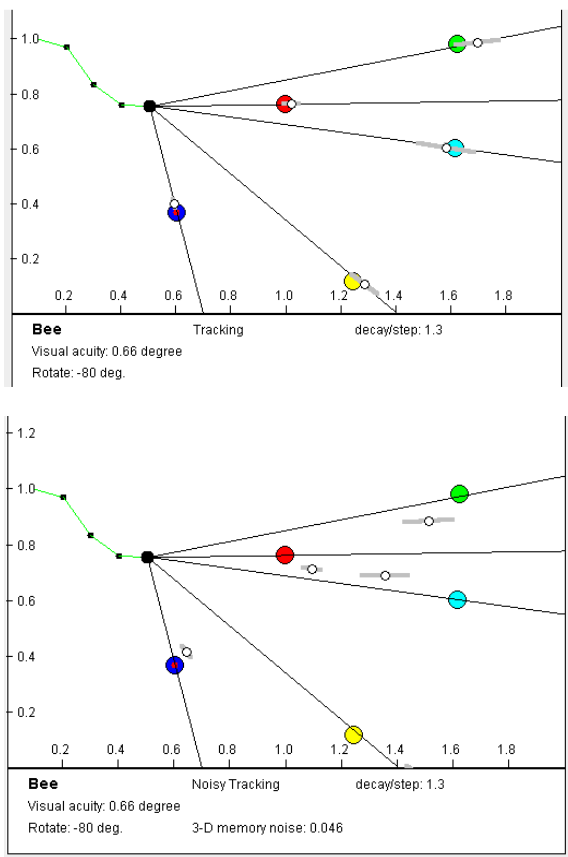

*Figure 6: The effect of random noise of 5% in memory storage of positions, on the accuracy of tracking. With noisy tracking, the estimated object positions differ significantly from the true object positions.*

Here, even by adding memory noise of under 5%, or one part in twenty (larger errors might be expected for storage of positions as neural firing rates) the estimated object positions have drifted considerably from their true positions in six steps, and the noisy tracked error bars are far from the true object positions – whereas without memory noise, tracking is reliable.

The same can be seen in range likelihood distributions like that in figure 5, and in figure 7 below – where Bayesian and Tracked estimates stay close to true object ranges, but noisy tracked estimates can be far from the truth:

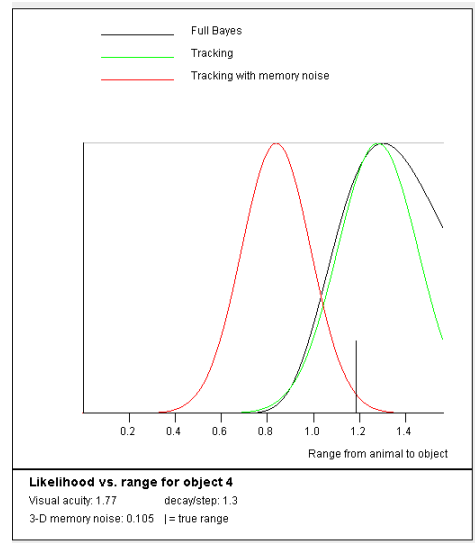

*Figure 7: Errors in estimating the range from the bee to one object, shown as likelihood curves for three types of estimation: full Bayesian (black), tracking (green) and tracking with memory nose (red). The true range of the object is shown as a vertical bar.*

Looking at individual estimates in these plots does not give a systematic overview. For this, it is useful to track many objects over several steps, with and without memory noise, to compare how both estimates deviate from true object positions. An example is shown in figure 8:

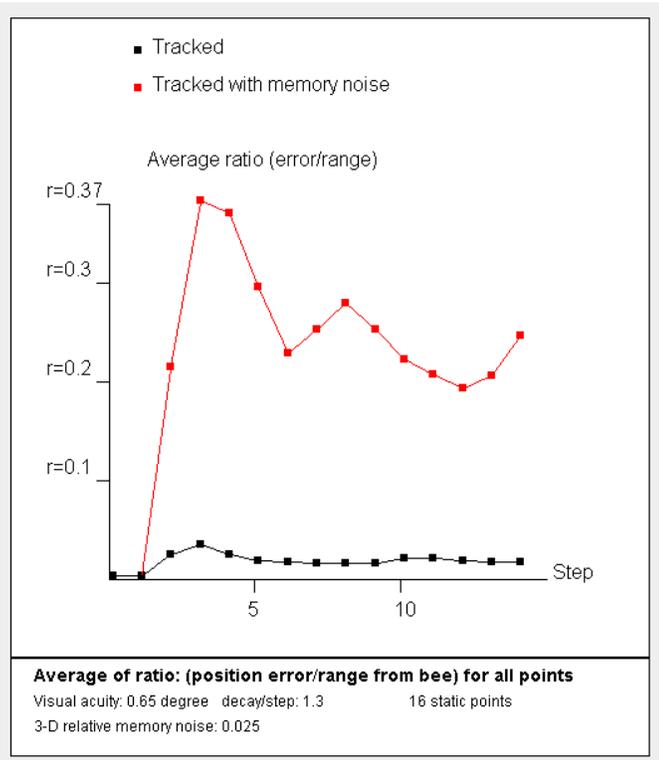

*Figure 8: Errors in position estimates depending on memory noise, plotted against number of steps in the bee's track. The black points show the tracking errors with no memory noise, while the red points show the effect of memory noise.*

Because errors in estimated positions are less important for distant objects, the graph shows a ratio, of (position error/object range)[4], averaged over all objects for each step. The red points are errors with memory noise, and the black points are errors without memory noise.

Even with a very low level of memory noise (2.5%, or one part in 40), the proportional errors with noise in this graph are of the order of 30%; while the proportional errors without noise are much lower, around 2%. Often tracking amplifies memory errors cumulatively, as the estimates drift away from reality. Generally the position errors in the presence of memory errors vary markedly between runs, and in some runs, errors can be much larger than the 30% shown here.

The drift of position estimates away from true object positions can be understood by an analogy with Brownian motion. At every step, each object's estimated position gets a random perturbation 'kick' as it is stored in memory. While these perturbations sometimes cancel between steps, they do not always cancel – and as in Brownian motion, the variance of the position from the true position grows with the number of steps[5]. Therefore over many steps, the divergence of the estimated position from the true position can grow, unless there is some restraining effect by retaining memory of the estimate from a few steps ago – which there is not in this model. Any such extra memory would complicate the model over the simple tracking model, and would increase its memory requirements.

The result of the model is that, while tracking with no memory noise can lead to a precise three-dimensional model of local space, even small random memory noise rapidly degrades the quality of the model.

An animal's internal model of local space needs to respond to changes very rapidly – within fractions of a second. Within such timescales, the precision of spatial memory from stochastic neural firing rates is low, typically with RMS errors greater than 10% - which is enough to degrade the tracking model – or any other model which relies on short term spatial memory (that is, any model capable of inferring three dimensional locations from motion).

The challenge of precise spatial memory is much harder for insects, whose vision is typically 5 times faster than ours. This gives them much less time for neural firing rates to build up a precise representation of a position. This is a serious challenge to neural models of spatial memory.

## 7. Motion Detection

An important use of an internal model of three-dimensional space is the detection of motion in other objects, when the animal itself is moving. Some motion detection cannot be done directly in the visual field, because objects apparently move in the visual field when the animal is moving, and as it makes visual saccades; so a 3-D model of space is required to detect true motion.

Detecting what is truly moving is very important, because moving things require immediate attention. For a predator, something moving might be a meal. For most animals, the reverse holds; if some predator has good motion detection, and if the animal fails to detect the predator and freeze or flee, it may become the meal. Increasing the efficiency of movement detection has been an evolutionary arms race between predators and prey, for half a billion years. We expect animals to do it very nearly as well as it can be done. Doing so requires the best possible 3-D model of local space.

Motion detection is important for another reason [Fields 2011]. If two points do not move relative to one another, they are perceived as parts of the same extended object. It is only relative motion that distinguishes objects from stationary background, and gives them boundaries [Fields 2014]. Objects with boundaries can be categorized; and from their categories, they can be expected to behave in certain ways.

---

[4] There is a low range cutoff, to avoid exaggerated contributions from objects which happen to be very close to the bee.

[5] Because of the exponential damping of past information, which is applied at each tracking step, this growth does not continue indefinitely.

The program can be used to test the efficiency of motion detection for bee vision, with tracking at different levels of memory noise. Just as Shape from Motion is only a part of visual perception, so the detection of motion modelled by the program is only a part of animals' detection of motion. In mammals, for instance, there are direct short cuts to detect possible motion very rapidly, before there is any time to update an internal model of space [Fields2011]. These fast methods, applicable mainly when the animal is not moving, are not modelled by the program.

To test motion detection in the program, you select one of a number of regular shapes, such as a hexagon or a cube. This shape is then inserted in the model at a random position, as a set of objects at its vertices. The shape is stationary for a defined number of steps. Then it starts to move, and continues to move with the same velocity for some steps.

The bee detects motion of the shape in a Bayesian manner. If the vertices of the shape are moving, then their fit to its initial tracking hypothesis (that objects do not move between steps) becomes worse; the negative log likelihood computed for each moving vertex, at its best fitted position, increases. The bee does not know which objects are vertices of the shape; it identifies possible moving vertices as those objects with the largest negative log likelihoods, and then allows the inferred vertices all to move by the same displacement, to improve their negative log likelihoods. If the net improvement in negative log likelihood is significant, the bee infers that the shape is moving, and moves the vertices by that displacement in its internal model of space.

In the program display, the moving shape leaves a red trail behind one of its vertices as it moves. An example is shown in figure 9:

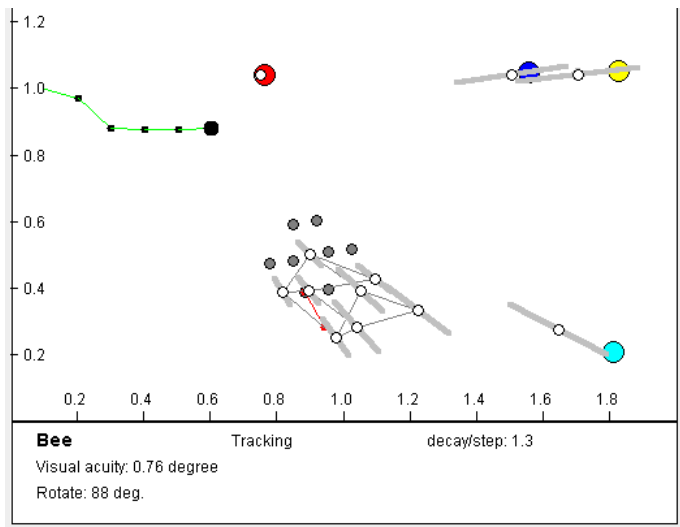

*Figure 9: The vertices of a cube are shown as small grey points. The cube has started to move, but the bee has not yet detected its motion – so the bee's estimates of the positions of the vertices (small white circles) lag behind the true positions.*

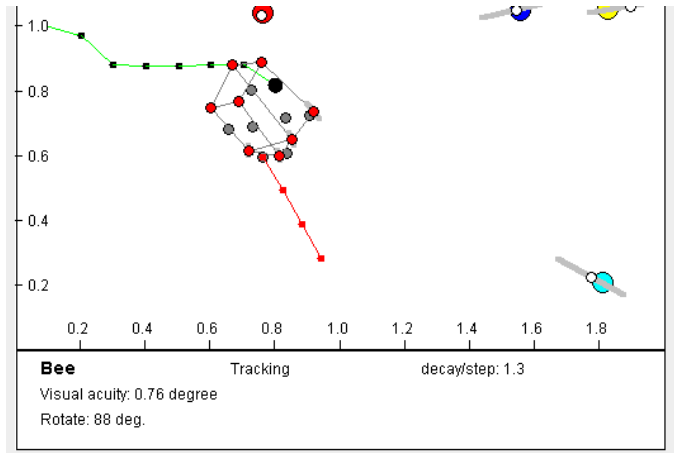

*Figure 10: the same run, two steps later. The cube has continued to move (small grey circles, with a red red trail of its movement). The bee has now detected motion, and its estimates of the positions of the vertices (now coloured red, to indicate motion) have caught up with the cube.*

Sometimes the bee fails to detect motion (e.g. if the shape is too far away from it). This can depend on several variable parameters, such as:

- The number of vertices of the shape
- The distance moved by the shape
- The distance of the shape from the bee
- The number of steps for which it moves
- The visual acuity of the bee

In this way, the bee has a certain probability of detecting motion, which depends on several variable parameters. Whatever the values of those parameters, the probability of detecting motion is degraded by adding memory noise to the tracking – which adds noise to the best estimated positions, making it harder to distinguish true motion from noise. This can be displayed in a graph, as shown in figure 9:

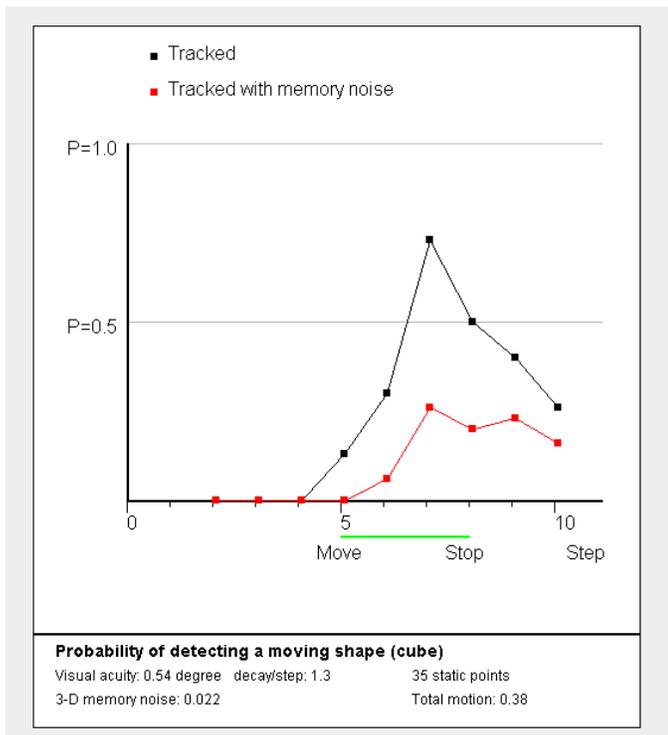

*Figure 11: Efficiency of motion detection is degraded by memory errors. As the object moves (green bar), a tracking bee identifies the moving shape correctly with high probability (black points). Memory noise less than one part in 40 degrades the efficiency of motion detection down to random levels (red points)*

Here, the green bar shows the steps in which the object moves. During these steps, a tracking animal has a high probability of correctly detecting the moving object. However, even adding a small amount of memory noise (here, 2.2%. one part in 45) degrades the efficiency of motion detection down to random levels.

The highest level of memory noise which allows effective motion detection appears to be about 1%.

This can be confirmed by running the program with many parameters varied (such as the amount of movement, visual acuity, level of memory noise, or the number of objects).

This is not a sophisticated model of motion detection, but it leads to an important result of the model: that the efficiency of motion detection is sharply degraded by even small amounts of memory noise.

Because of the strong and sustained selection pressure for the best movement detection, we expect animals to do it very well (which they do, as can be confirmed by observation or experiment). This poses a severe problem for neural models of spatial short-term memory, because of their high expected level of memory errors.

## 8. Multi-Sensory Integration: An Aggregator Model

For vision, the tracking model works by:

- Storing a best estimate position vector for each object, together with its uncertainty tensor (in a Gaussian approximation, where the negative log likelihood is quadratic in position)
- To update the object's position at each step, multiply the previous probability density function by a near-Gaussian cone along the line of sight, from the latest sighting. This is done in the model program by adding the negative log likelihoods.
- Find the maximum of the new probability density

This produces a near-optimal model of 3-D positions from simple vision and the bee's motion, using the Bayesian prior probability that most objects do not move as the bee moves.

The fact that tracking works with two very different types of sensory input (vision and echo-location) suggests that it can easily be generalized to many other types of sensory input.

The tracking model can be generalized to multi-sensory integration. If there are other sensory constraints on the location of an object, their probability density can be multiplied by the probability densities from the previous tracking estimate and from vision. Negative log likelihoods from all knowledge sources can be added, or aggregated [Worden 2020b]. This finds the maximum of the Bayesian posterior probability density, in the light of all sense data. It is close to the Bayesian optimal form of multi-sensory integration. It can be applied to integrate many different knowledge sources, including:

- Stereopsis
- Proprioception (limb positions)
- Sound
- Shape from shading
- Occlusion
- Object recognition (which constrains positions of points in objects to fit their shapes)
- Expected object dynamics

Olfaction and taste are less useful in this regard, and contribute little.

All animals need to use multi-sensory integration, to build the best possible 3-D model of space, reflecting the consequences of their actions. The shape from motion model of this paper could be a foundation for how they do it.

## 9. Neural Spatial Memory and Cognition

An internal 3-dimensional model of local space plays a central role in cognition, being used at many moments of the day for many diverse purposes such as the control of physical movement, recognition of objects in space, manipulating objects, and detecting moving things.

This model of space is built from sense data which typically (like vision) is two dimensional. A core technique for doing this (as modelled in this paper) is to infer object positions from the animal's own motion, using Structure from Motion. The results of these models are:

1. A tracking algorithm can build a good 3-D spatial model from vision – almost as good as the best possible model that can be built.
2. The tracking model requires a short-term memory for the 3-D locations of things, to compare the animal's current view with recent views.
3. If the spatial memory has even modest levels of noise, it rapidly degrades the quality of the model and the effectiveness of motion detection.

How could this model be neurally implemented in brains? Points (1) – (3) above pose problems for neural models of cognition. Thise problems are discussed in this section.

The first question is how to represent three-dimensional information by neural firing rates, as is required for short-term neural spatial memory. I consider three options:

A. Represent two of the dimensions by position in a 2-D neural sheet, and represent the third dimension by firing rates at locations in the sheet – like a visual cortex with depth
B. Represent all three dimensions of any position by neural firing rates
C. Represent all three dimensions by positions of neurons in a 3-D 'clump' of neurons.

Option (C) is unattractive for several reasons. To give high spatial resolution, it would require to be a prominent clump of neurons, which has not been observed in brains, and would pose serious problems of neural connectivity, for neurons in the middle of the 3-D clump.

Option (A) potentially gives high resolution in two of the dimensions, represented by positions in the sheet. A possible drawback is that for motion detection, it would be useful to represent space in an allocentric frame of reference, where most things do not move; but the visual cortex does not use such a frame, so the spatial memory would need to be somewhere else in the brain.

Option B raises questions about how the three dimensions are defined (in what kind of coordinate system?); and then, how the commonly needed operations, such as vector addition of displacements, would be computed. It is hard to devise any simple representation of three-dimensional space by neural firing rates, and then to use it to do spatial computations simply. This issue also arises in option (A); there is little to say about it, except that complex forms of processing appear to be needed, and that these have not yet been devised.

Most important is the issue raised by point (3) – that small levels of memory noise degrade the quality of the model and the effectiveness of motion detection. In the model, memory noise as little as 1 part in 40 degrades motion detection down to near-random levels. How could a neural spatial memory give better precision than that?

Quantities like components of vectors can be represented by stochastic neural firing rates. If a single neuron fires stochastically with N action potentials per second, in one second it represents information with precision approximately one part in $\sqrt{N}$. To get a precision of the order of 1% (as appears to be required to support motion detection) in one second would require $N = 10,000$, which is an unrealistically high firing rate. Typical neural firing rates are 5- 50 pulses per second.

The problem is more serious because animal brains need to choose physical actions faster than once per second. For a small mammal, the required times may be of the order of 100 milliseconds. For insects, whose vision is 5 times faster than our own [Chittka 2022], the timescales may be a few tens of milliseconds. It is only possible to fit a small number of neural firings into this time – giving very poor precision.

To represent spatial coordinates by stochastic single neuron firing rates requires an impossible tradeoff between speed and precision. It cannot be done.

We must look for other solutions. One possible approach is parallelism. Using the numbers above, a very high degree of parallelism would be required to get the required aggregate firing rate – perhaps 100 parallel neurons to represent one dimension of one position (as the precision increases with the square root of the number of parallel channels). As we know that brains represent the positions of many objects at once, this soon scales up to prohibitive numbers of neurons – particularly in insect brains, where there are fewer than a million neurons in total. Another approach would be to use some non-stochastic firing pattern, such as regular bursts. Another approach would be to use a more complex encoding of distances, perhaps using small linked groups of neurons to encode coordinates in multiple firing rates.

A drawback of all these approaches is that they make an already complex problem – how to do spatial computations, such as vector additions – yet more complex. Some of the possible approaches would cause distinctive neural connectivity or firing patterns, which might be looked for.

This discussion has only sketched some of the issues that need to be addressed. The main conclusion is that current models of neural information processing in the brain cannot address the core problem of spatial cognition, because of the very difficult tradeoff between speed and precision. Some new ingredient is required.

Precise spatial cognition is a central requirement for all animal brains; but apparently, neurons cannot do it. It is a high priority to explore neural implementations of 3-D spatial cognition, to see if there is any possible solution.

# 10. Active Inference Models of Spatial Cognition

The Free Energy Principle with Active Inference (FEP/AI) [Friston Kilner & Harrison 2006; Friston 2010] is an appropriate framework for modelling 3-D spatial cognition, for several reasons:

- It has a well-defined formal framework, which addresses perception, action, and cognition.
- It is applicable to any domain of cognition
- It is built on a Bayesian foundation, so it is well suited for computing near-optimal Bayesian spatial cognition
- It has a defined neural process model, which has been tested in many applications.

Here I briefly review some relevant papers on FEP/AI.

'A step-by-step tutorial on active inference and its application to empirical data' [Smith, Friston and Whyte 2022] describes the mathematical formalism of Active Inference, its Bayesian foundations, and its neural process model (neural implementations). Section 5 describes the neural process model. It describes the meaning of neural diagrams like figure 12 below (taken from their figure 12), which frequently appear in Active Inference publications.

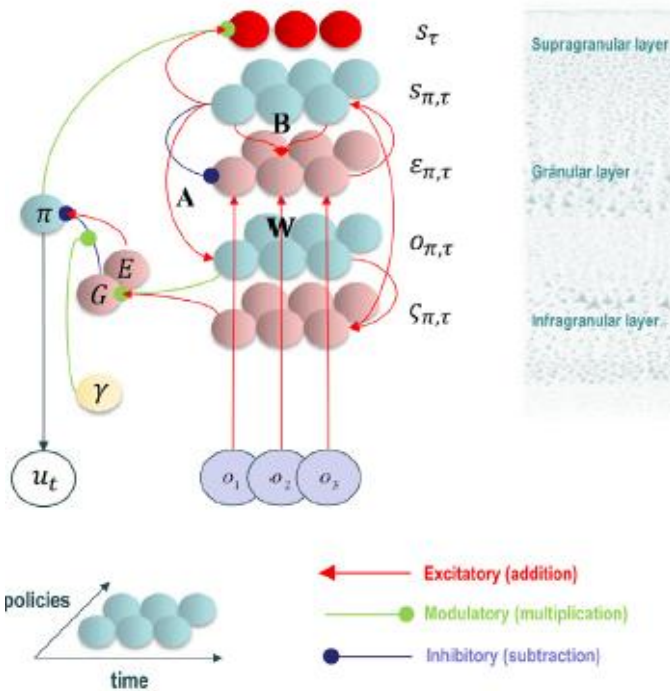

*Figure 12: neural process model diagram, from [Smith,Whyte & Friston 2022]*

The spheres represent neurons or groups of neurons (here shown in a cortical column) and the lines represent neural connections. The models assume that quantities (such as Bayesian probabilities) are represented by neural firing rates, and that a group of neurons can carry out the operations of addition, subtraction, and multiplication. These operations implement the mathematics of active inference. Prior probabilities are generally represented by synaptic coupling strengths.

The tutorial does not discuss the precision of neural representation of numerical quantities. It appears that the MATLAB active inference toolkit, referred to in the tutorial, does not model the errors in neural storage of quantities.

If the firing rates are stochastic, and there are N firings in some short timescale T, then the precision of representing some quantity is approximately one part in $\sqrt{N}$ – which would be insufficient precision to support the models of this paper. Some different encoding of positions would be required to give the required precision; that would complicate the neural computations such as multiplication or addition.

While there are Active Inference models of spatial cognition, I do not know of any models which explicitly address three-dimensional spatial cognition in animals, or which address the issue of the required precision in neural representation.

When active inference is applied as Active Vision [Parr, Friston et al 2021], it uses a hierarchical model of three-dimensional space, in which some of the quantities represented by neural firing rates are the components of relative displacements between objects (high in the hierarchy) or parts of objects (lower in the hierarchy).

The Active Vision work closest to the concerns of this paper is that by [Van Der Maele et al, 2023a, 2023b] on robotic understanding of 3-D scenes, where a robot can move a camera in three dimensions to understand a 3-D scene. Here, active inference has been shown to give better performance than passive learners. Even in this work, there are important differences from the use of active vision in a biological context; for instance, in robotic applications, the issue of neural imprecision does not arise.

'Deep Active Inference and Scene Construction' [2020] describes how agents infer a higher-order visual pattern (a "scene") by sequentially sampling ambiguous cues. While it has some relation to the current work, its conclusions are mainly qualitative, and it does not address three-dimensional modelling of space. The main computational model concerns the analysis of random dot motion in two dimensions.

Three-dimensional spatial cognition is fundamental for animal brains, and is a challenging problem for neuroscience. The time may be right for the FEP to be applied to this challenge.

# 11. Wave Storage of Spatial Information

This section describes an alternative to a purely neural implementation of spatial cognition.

For any computation, if the physical computing mechanism matches the physics of what is computed, the computation can be simpler, more economical and more precise. It is possible to 'let the physics do the computing' – directly, rather than using some complex device to compute indirectly.

This principle was responsible for the early uses of analogue computers, before digital electronics became predominant.

The same principle might be applied in spatial cognition. If there is some approximately spherical volume in the brain, which can hold wave excitations, then each wave can have a different wave vector, or **k**-vector. This is a three-dimensional vector which describes both a wavelength and a direction of wave motion (orthogonal to the wave front). If the physics of the wave is linear, the same volume can hold many independent waves, with different **k** vectors – which do not interfere with one another.

Therefore a wave can be used to store the independent locations of many objects – objects with different locations **r**, related to the wave vectors by **k**=α**r**, where α is a constant. If we assume that:

- Each wave excitation can persist for short periods (fractions of a second)
- The minimum possible wavelength λ is small compared to the diameter D of the volume (so there is a large range of possible **k**-vectors, in all directions)
- Neurons can couple selectively to waves of different wavelength and direction, as transmitters and receivers (e.g. one neuron might have its wave receptors or transmitters aligned and spaced with the wave fronts, to be selective near one **k**-vector) then neurons could use the wave as a short-term memory for the locations of many objects. This form of spatial memory could have major benefits over storage in neural firing rates:

1. The three dimensions of the wave correspond directly to the three dimensions of object positions; there is no need for any preferred direction, or for the representation to be asymmetric between directions. There is no need to choose a coordinate system. It is a simple and natural representation of positions.
2. A large number of **k**-vectors (independent object positions) - of the order of $(D/\lambda)^3$ - can be stored in the same wave volume.
3. The precision of each object location in any dimension is approximately one part in $(D/\lambda)$ – which can be better than one part in 100, as appears to be required for effective motion detection.
4. As in a hologram (which works by the same principle) there is very little spatial distortion of positions.
5. In principle, the wave can be updated very fast, say within a few milliseconds. There is no hard tradeoff between speed and precision.

These are major benefits, possibly overcoming the serious problems with neural storage described in the previous section. They are enough to make the tracking model of spatial cognition workable – which it seems not to be, with purely neural storage of positions.

There are other possible benefits, if we make further assumptions about the wave and how neurons couple to it. Further potential benefits are:

6. A single neuron's coupling to the wave might be tunable to different wave vectors – if the sensitivities of its wave transducers (possibly synapses) can be altered, or can be given phase delays, by a steering signal. This opens up the possibility for the wave to be used for selective **spatial steering** of signals – something which is needed for dynamic routing of information, for instance from sense organs to specialized pattern recognition modules.
7. Individual neurons could be tuned not just to specific wave vectors (represented positions) but also 'de-tuned' to represent the uncertainty in object positions, as a Gaussian-like spread of wave vectors. Representing uncertainty tensors is required by tracking model.
8. If the phases of the waves could be controlled, physical addition of waves from different sources might directly represent the addition of negative log likelihoods (or free energies [Friston 2010]). This is required to find Bayesian maximum likelihoods, and in the aggregator model, for multi-sensory integration.
9. If the wave has several internal degrees of freedom (as, for instance, polarization of an electromagnetic wave gives two degrees of freedom), these could represent different attributes of objects, such as colour.

If spatial positions are stored in a wave in the brain, there must be some minimum possible wavelength $\lambda_{min}$ that neurons can couple to; which implies that there is a maximum **k**-vector, and a maximum distance that can be represented. This is a problem for representing very large distances, which animals sometimes need to do. So the wave excitation cannot represent Euclidean space directly, but probably represents some transform of Euclidean space, designed to minimise geometric distortions at small distances. In this respect, projective transforms of space [Rudrauf et al 2017, 2022] are particularly useful, as they preserve straight lines; and straight lines are important for controlling motion and recognizing shapes. So the wave storage may use some near-projective transform of Euclidean space.

This may be why we see the stars as a spherical canopy, and why perception has minor distortions of Euclidean geometry, especially at large distances.

Storage in a wave has potential for a spatial short-term memory, which could be greatly superior to neural memory. Many details remain to be worked out – including:

a) What is the physical nature of the wave?
b) How can the wave be sustained for the necessary times?
c) What is the source of energy for the wave?
d) How do neurons couple to it? What genes and proteins are involved?
e) Can neurons have steerable coupling to the wave?
f) Where in the brain does the wave reside?

There are candidate answers for (f), described in the next section.

## 12. Evidence for Wave Storage of Spatial Information in Brains

Related papers [Worden 2020a, 2024b] propose that:

- There may be a wave excitation in the mammalian thalamus, storing spatial information
- The central body of the insect brain may play the same role

Evidence supporting the hypotheses is described in the papers. Here I summarise the most important evidence, common to the mammalian thalamus and the insect central body:

1. Both the thalamus and the central body have a simple, near-spherical anatomical form, which is remarkably well preserved across a wide range of species [Sherman & Guillery 2006; Heize et al. 2023]. A near-spherical shape is well suited to hold a wave in three dimensions. This shape is in marked contrast to the contorted, variable and species-specific shapes of most parts of the brain (such as the cortex, the hippocampus, or the mushroom bodies). This strongly suggests that in the thalamus and the insect central body (and not in other parts of the brains), something other than neural synaptic connections is going on, which could be a wave.
2. Both the thalamus and the central body are centrally located, and richly connected to other parts of the brain – suggesting that they both act as an integrating hub role for some important type of cognition. Spatial cognition is such a function, because it underpins everything an animal does in its lifetime.
3. Both the thalamus and the central body are innervated by sense data of every modality – except, remarkably, olfaction[6]. Smell is of little use for the precise, fast, location of things in a 3-D model of space. This may explain why smell is not linked with those central brain parts.
4. Both the thalamus and the central body are closely linked to consciousness [Baars 1988, Chittka 2022]. Lesions in the thalamus, or targeted predator injection attacks on the insect central body, cause cessation of consciousness.

For the mammalian thalamus, there is a further strong piece of evidence. Without a wave, the anatomy of the thalamus does not make sense. The thalamus is a cluster of centrally-placed thalamic nuclei, with few or no connections between nuclei [Sherman & Guillery 2006]. In terms of brain energy consumption and neural connections, this does not make sense. The same neural connectivity (i.e. the same neural computation) could be obtained with less brain energy consumption (less net axon length) if the thalamic nuclei separated, migrating outwards towards the cortex. The anatomy of the thalamus only makes sense if the thalamic nuclei need to stay together, as they would need to do, to be immersed in the same wave.

Taken together, this evidence is very significant. The wave hypothesis accounts for striking neuro-anatomical facts, which are not explained in a purely neural model of the brain. This is sufficient reason to justify further exploration of the wave hypothesis.

## 13. The Wave Hypothesis and Theories of Consciousness

Leading theories of phenomenal consciousness such as [Baars 1998, Tononi 2012] propose that consciousness is linked to neural activity in the brain. I suggest that these theories are currently at an *impasse*, because of the spatial, geometric nature of conscious experience. Consciousness is largely consciousness of the space around us, and most neural theories do not account for that:

- Most theories of consciousness suggest that it is caused by some aspect of neural activity in the brain, such as free energy minimization [Solms 2019] or a global workspace [Baars 1988] or integrated information [Tononi 2012]. These may be possible accounts of why consciousness **exists**

---

[6] Olfaction, unlike all other sense data, does not pass through the thalamus on its way to the cortex. The insect central body has no direct connections to the mushroom bodies, centres of olfactory learning. However, it may have other connections to olfaction.

– but they say little about the **properties** of consciousness, particularly its spatial properties. A notable exception to this is the Projective Consciousness Model (PCM) of [Rudrauf et al 2017, 2022], which does address the spatial form of consciousness. These spatial geometric properties are the only rich source of data with which to test a theory of consciousness – the only way to get strong confirmation for a theory.

- If a neural theory is to account for the spatial form of consciousness, it needs to account for the fact that the form of consciousness is remarkably like the form of real space around us. That is 'what it is like' to be us. All known neural encodings of spatial information in the brain are highly distorted, in species-specific ways. So any neural theory of consciousness needs to define how these complex, variable, neural encodings of space are decoded, to give our undistorted experience of space. No neural theory of consciousness attempts this, and it seems to be a hopeless endeavour – because a complex decoding have many variable parameters, making any neural theory of consciousness too complex to test.

The wave hypothesis of spatial memory offers a way past this neural *impasse*. Consciousness may not be related at all to neural activity, but arises from the wave excitation in the brain. This would give a simple account of the main empirical fact about consciousness – that the form of consciousness is very like real space. The form of the wave is very like real space – so if conscious experience arises directly from the wave, it will be spatially undistorted, as we experience it. This gives a very good, unforced fit between theory and data. As the wave in the brain probably holds a projective-like transform of Euclidean space, the resulting theory of consciousness is a wave realisation of the PCM of [Rudrauf et al 2017, 2022]. It is explored in related papers [Worden 1999, 2024c].

## 14. Does Fitness Beat Truth?

In a series of papers on the Interface Theory of Perception, [Hoffman 2009, 2017; Hoffman, Singh & Prakash 2015]. have proposed that the purpose of perception is not to build veridical internal models of the external world, but only to maximise fitness; and that therefore, perception is not veridical. That proposal runs directly counter to the proposals of this paper; so it needs to be addressed.

The papers by Hoffman et al. describe and cite models of evolutionary games and genetic algorithms, to conclude that animals which use Bayesian maximum likelihood models of their surroundings cannot compete with 'fitness-first' animals which simply choose actions to maximise fitness; so that veridical Bayesian internal models are not used in nature.

A paper by [Prakash, Stephens, Hoffman, Singh and Fields 2021], 'Fitness Beats Truth in the Evolution of Perception' contains a proof, in an evolutionary game-theoretic model, that a 'fitness-first' animal, which only uses its sense data to maximise its fitness payoff in any encounter, will always out-compete a 'truth-first' animal, which uses sense data to build veridical models, by Bayesian maximum likelihood. These two species are pitted against one another in a series of competitive encounters. The paper shows that the fitness-first animal always drives the truth-first animal to extinction; that Fitness Beats Truth (FBT). The authors conclude that all species are fitness-first, rather than truth-first.

They say that the FBT result applies to visual perception, casting doubt on any research in vision [e.g. Knill & Pouget 2004] which uses Bayesian veridical scene reconstruction. This paper has described the Bayesian construction of a 3-D model of local space from vision or echo-location; it is a veridical model, being geometrically consistent with real space. If the FBT result applied to all vision, it would cast doubt on the results of this paper. So we need to understand the relation between the results.

In [Worden 2024a] I showed that the cognition with the greatest possible fitness is to choose actions 'as if' by computing a Requirement Equation (equations 11 and 16 of that paper) which resembles Bayes' Theorem, but is not the same as it. In Hoffman et al.'s terminology, the requirement equation is what a fitness-first animal computes (or acts as if it had computed; sometimes short cuts are possible).

I showed that in some cases, to choose actions in this way requires an animal to construct internal models of reality using Bayes' Theorem; but in other cases, it does not require an internal model of reality. The cases are distinguished as follows:

a) Internal cognitive models are required where there are many choices of action, or complex choices of action, depending on the same aspects of the external state of affairs; in brief, where the choices of action have high information content.

b) If the choices of action do not have high information content, then it may not be necessary to build a Bayesian internal model – because there may be a short-cut or 'as if' computation, which gives the fittest choices of action, with less effort.

The animal's brain is an information channel from external reality, through its sense data, to its choices of action. If those choices have high information content, it is worth building an information-rich Bayesian model of reality, as a high-capacity channel; but if the choices have low information content, that may not be worth the effort, as simpler short cuts may suffice.

In the case of vision, many complex choices of action depend on the locations and motion of external objects. Some important types of spatially-dependent choices are:

1. Planning and executing one's own movements – controlling limbs
2. Detecting and attending to moving objects, while moving
3. Manipulating objects – grasping, eating, throwing
4. Recognising things and shapes – straight lines, bends, plane surfaces.
5. Anticipating how things will move – e.g. how they will fall, or carry on moving in a straight line.

The choices in (1) and (3) are complex sequences of graded muscular forces. The summed information content of these decisions, which all depend on the geometry of real external space, is very large. For these purposes, the internal representation of that geometry needs to be veridical, obeying the same constraints as real space – straight lines in real space need to be nearly straight lines in the model, and so on.

For instance, there is no geometric distortion of the spatial model which could make motion detection (2) more efficient; but any geometric distortion would degrade performance in the other categories, particularly (1) and (3).

I conclude that visual perception is of type (a), where veridical internal models of reality are required. Here, fitness needs truth.

How should we understand Hoffman et al.'s models, and their FBT result? Are the models of type (a), where fitness needs truth, or type (b), where short-cut cognition suffices?

The core assumptions of the FBT model and the assumptions of the Requirement Equation [Worden 2024a] are mutually consistent. In both models, animals choose actions to maximise their Darwinian fitness, in a series of encounters though their lives; and this leads to changes in phenotypes over many generations.

In the FBT proof, a fitness-first animal meets a truth-first animal in a series of competitive encounters, where:

1) Both animals can see the same set of 'territories', and both receive the same sense data, which gives them the same partial information about 'resources' (such as food) in the territories
2) The truth-first animal constructs a Bayesian maximum likelihood model of the resources in each territory. Its first choice of territory in the territory which has nearest to its required amount of resources in its most likely model.
3) The fitness-first animal chooses the territory which maximises the average fitness payoff from resources it will get, in all models weighted by their probability.
4) Based on a 50% chance, one of the animals can go to its first choice of territory. If the other animal has made the same choice, it has to go to another territory - its second choice.
5) Each animal has to consume all the resources in the territory it goes to (which may be more than it needs); the relation between resources and fitness is not monotonic.

The differences between truth-first (2) and fitness-first (3) are subtle, but they are different computations. Truth-first looks at maxima of a likelihood distribution, while fitness-first integrates over likelihood distributions to compute a mean fitness. The requirement equation of [Worden 2024a] is a fitness-first choice.

The key point to note about Hoffmann et al.'s model is that each animal has only one simple choice of action – its preferred choice of territory. This choice has small information content. No Bayesian internal model of reality is required; some simpler short cut can do the job. It is not so much that fitness beats truth; for simple choices of action, fitness does not need truth.

In another paper 'Fact, fiction and fitness) [Prakash et al 2020), the same authors discuss evolutionary games in four mathematically defined 'mini-worlds', designed to emulate aspects of the animal world. The mini-worlds are:

1. Total orders: to emulate graduated perceptions of magnitudes
2. Permutation groups: to emulate rearrangements of objects
3. Cyclic groups: to emulate properties of spacetime
4. Measurable spaces: to emulate probability distributions.

In these worlds (W), they equate veridicality of perception (P) with homomorphic mappings W=>P. These are monotonic functions, which do not 'scramble' percepts. In each mini-world they show that the mappings W=>V from the world to fitness payoffs (V) are not homomorphic mappings, with very high probability. They conclude that the mappings W=>P cannot be veridical.

In this analysis, the choices of actions (A) are scarcely mentioned. However, it appears that the choices of action are simple (like eat/don't eat' as in their 'critrs' example), and that all four mini-worlds lack the key feature that makes veridicality necessary: the existence of many information-rich choices of action, all depending on the same model M of reality. In other words, no Bayesian model of reality is required, and short-cut cognition will suffice to choose actions. So the four models do not apply to vision.

The result that Fitness Beats Truth only holds in models like those of [Prakash et al., 2020, 2021] where the choice of possible actions is simple. In more complex cases, where many different action choices all depend on the same aspect of reality, cognition requires an internal model. I showed in [Worden 2024a] that these models are constrained to be veridical[7]. The result that Fitness Beats Truth does not apply to visual perception, or to the 3-D model of space derived from it. For that model, fitness needs truth.

Vision is used to build three-dimensional models of local space, which have two main classes of applications:

1. The control of physical movements
2. Categorisation, for discrete choices of action

The Interface Theory of Perception, and the FBT Theorem, address only applications of class (2). They do not address decisions of class (1), which require veridical models of spacetime, at all moments of the day.

Within class (2), some kinds of categorization still require veridical models. We can place them on a spectrum from (2a) to (2b):

(2a) **Principled Categorisation**: where the categories depend on physical properties which can only be computed from a veridical spatial model.

(2b) **Pattern-Based Categorisation**: where the category may depend on sense data in complex ways, requiring pattern recognition rather than computation.

This distinction is not binary; there is a spectrum, and the ends of the spectrum can be illustrated by examples.

Movement detection (as modelled in the program) is at the principled end (2a). To decide: 'Is that object moving or not moving?' the program uses a veridical model of spacetime. However, animal motion detection uses short-cut patterns as well [Fields 2011,2014]; so motion detection is a hybrid.

The examples cited in the Interface Theory of Perception, and the FBT Theorem, are at the (2b) end of the spectrum. Often the categorization "Is this situation of type A or type B?" means in effect: "Should I do A or B?" – as in the jewel beetle's choice: "Should I try to mate?". Neural nets do this kind of categorisation well, and it is a form of short-cut perception; neural nets use no internal models of reality[8].

In a further complexity, many categorisation choices are learned, rather than innate. The choice "is that fruit ripe enough to eat?" may need to be learned. This applies to choices dependent on features of the habitat which change too fast for the evolution of brains to catch up.

Along the spectrum (2a)-(2b), the evolutionary choice of whether to use short-cut cognition is a fitness tradeoff: if the fitness costs of short-cut cognition are small enough, and if it saves brain energy costs, it may persist indefinitely (e.g. if there are so few beer bottles in the wild, that the jewel beetle can use short-cut perception to identify a mate; or if predators are so rare that it is not worth the cognitive cost of seeing through their deceptions)

The Interface Theory of Cognition and the FBT Theorem apply to pattern-based categorization, where the patterns are perceived like computer icons. For the other uses of vision, truth is needed.

We do not perceive all of space veridically. In the space of our conscious experience, there is no boundary between the outer space of the world we see, and the inner space of our bodily feelings [Fields 2014, 2023]. External things, limbs, and internal feelings are all in the same conscious space, with no boundaries. We perceive outer space and limbs veridically, because we need to move skillfully. We perceive the inner space non-veridically, as a set of icons or emojis for social situations – particularly influencing our self-esteem [Worden 2024d]. This non-veridicality has had enormous consequences for mankind.

## 15. Conclusions and Further Work

Animals need to build internal models of the 3-D space around them, in order to control their every movement, and for other reasons. The best way to do this is to build a Bayesian maximum likelihood model of 3-D space, using sense data. Any other model of space (or no internal model at all) would be less fit; such an animal would be driven to extinction by fitter species.

There has been massive selection pressure to make the internal model of space as precise as possible, and animals appear to do it very well. This leads to a **Spatial Modelling Hypothesis** – that animals' internal models of 3-D space are very close to the best possible model, given their sense data.

This paper has described a program which computes the best possible spatial model, by direct use of Bayes' Theorem. It does so for bees – using vision of limited resolution – and for bats, using echo-location. In both cases, it computes object locations from the animal's own movement, using the strong Bayesian prior probability that other objects do not move as the animal moves.

The results of the Bayesian optimal model computation are:

- The Bayesian computation results in a faithful 3-D model of external space

---

[7] In [Worden 2024a] I showed that the best Bayesian model to use for multiple choices of action is not exactly a maximum likelihood Bayesian model, but is a hybrid model which selects the mean values of state variables within maximum likelihood peaks. This does not alter the point that the brain needs to construct near-veridical internal models for maximum fitness. They are hybrid models, not exactly maximum likelihood models.

[8] Large Language Models (LLM) suffer from a form of short-cut learning [Du et al 2024], when they succeed in tests by latching on to superficial properties of the prompts. In some sense, LLMs always use short cuts; they have no veridical model of the world.

- It requires retaining sense data from several previous steps of the animal's track, to compute 3-D object locations from the animal's motion
- Sense data need to be retained and retrieved with high precision. The precision and capacity are very demanding in short-term memory.
- If objects do not move, the spatial precision of the model after several steps can be better than the spatial precision of the sense data in each step.

The program computes a tracking approximation to the Bayesian model, in which sense data are not retained, but where sense data from the current step are compared with estimated object locations from the previous step. The results of the tracking model are:

- The tracking computation, like the Bayesian computation, produces a high-fidelity 3-D model of external space
- Less spatial short-term memory is required for tracking; however, estimated object locations must be stored with high spatial precision.
- Tracking requires storage of tensors representing the uncertainty in the position estimates.
- If estimates are stored and retrieved with high spatial precision, the model is very nearly as good as the optimal Bayesian model.
- Because it requires less memory and simpler computation, tracking is more feasible to implement in animal brains.
- A key use of the internal 3-D model is to detect motion of other objects, while the animal moves. This cannot be done directly from visual data, but it can be done from the 3-D model.

The program also computes a tracking model with random memory errors, of four possible types. The results of this model are:

- Even small random errors in the storage and retrieval of estimated object locations sharply degrade the performance of the model.
- This result is largely independent of the type of memory error
- The noisy tracking model degrades as estimated object locations drift away from their true locations, in a manner like Brownian motion.
- Small random errors in spatial memory sharply degrade the performance of motion detection.

The program is available and can be run with many variable parameters, to explore these findings.

I described how the visual tracking model can be generalized to multi-sensory integration.

The main result of this paper is that even small random errors in spatial memory, of the kind which are expected if short term memory is stored in neural firing rates, sharply degrade the precision of the model, and degrade its performance in core tasks like motion detection. For neural spatial memory, there seems to be an impossible tradeoff between speed and precision.

Given the central role of the spatial model in many cognitive tasks, including movement, which is tested at every moment of the day, this is a core challenge for cognitive neuroscience. It cannot be avoided just because it is hard. Without a neural model of spatial cognition, neuroscience might be compared to a theory of planetary motion without a sun; or an atom without a nucleus. The challenge of building a neural model of spatial cognition might be addressed in the framework of the Free Energy Principle, or other neural processing frameworks.

I described an alternative to neural memory, in which spatial short-term memory is stored in a wave excitation in the mammalian thalamus, or in the insect central body. I described the possible benefits of wave storage over neural storage, and summarized the evidence for a wave excitation in insect and mammalian brains. The wave excitation could be the basis of a theory of consciousness.

In parallel with efforts to find a neural model of spatial cognition, the wave hypothesis can be explored – in mammals, in insects, and possibly in other species, even in single-celled organisms. Compared to orthodox neural models of cognition, which are now well-trodden paths of research, exploring the wave hypothesis is green-fields research, where new ideas are needed and discoveries are to be made. If the wave hypothesis was confirmed, it would profoundly impact all areas of neuroscience.

Irrespective of the wave hypothesis, there are fruitful investigations to be made of the spatial modelling hypothesis – investigating the precision of the internal 3-D model in many different species, to see how close they come to the best possible model. For this, motion detection is a possible task. For instance, bees could be trained to find food on moving artificial flowers, and the efficiency of their movement detection tested. Does it come close the best detection possible, given their visual acuity?

I described how results of this paper are not consistent with the Interface Theory of Perception of Hoffman et al., where visual perception is concerned. Their results (and their proof that 'Fitness Beats Truth') only apply to simple choices of action, and to categorisation – not to the many complex choices of action which depend on vision.

## Appendix: Using the Program

The program used in this paper can be downloaded from www.bayeslanguage.org/bb/BB.zip Unzip the file BB.zip for additional material to this paper, including the runnable model program BeesAndBats.jar.

If you already use Java, the model program can be started by double-clicking the jar file, and following the instructions in the Help menu. An extract from the Help is:

**Getting Started**

(1) Press 'Start', then 'Run'.
The bee moves across the space. As it moves, small circles with error bars show the estimated point positions, in its internal 3-D model of space. Lines of sight from the bee to the points are shown

(2) Press 'Restart', then press 'Step' repeatedly.
The bee moves along the same track, in slow motion. See how the estimates converge to actual positions, and error bars get smaller.

(3) Use the 'view rotate' slider.
This shows the space from different angles, to see the third dimension.

(4) Press 'Start', choose 'cube' from 'Moving Shapes', then press 'Run'.
The corners of a cube appear. After a few steps, the cube starts to move. If the bee detects that the cube has moved, the corners of the cube turn red.

If you do not have Java, you need to download and install a Java Runtime Environment (JRE), for instance from https://www.oracle.com/uk/java/technologies/downloads/. This is an automated installation, taking about a minute.

The zip file also contains a folder of the program source code. You can import this into a development environment such as Eclipse or IntelliJ, to modify or extend the program.

## References

Baars, B. J. (1988). A Cognitive Theory of Consciousness. New York, NY: Cambridge University Press.

Chittka, L. (2022) The Mind of a Bee, Princeton University Press, Princeton, NJ

Cox, R. T. (1961). The algebra of probable inference. London: Oxford University Press

Du M., He F, Zou N., Tao D., and Hu X. (2024) Shortcut learning of large language models in Natural Language Understanding', Communications of the ACM, Vol 67 no 1.

Fields C. (2011) "Trajectory recognition as the basis for object individuation: A functional model of object file instantiation and object token encoding," Frontiers in Psychology 2, 49

Fields, C. (2014) This boundary-less world. In: D. Chopra (Ed.) Brain, Mind, Cosmos: The Nature of Our Existence and the Universe. Carlsbad, CA: Chopra Foundation. Ch. 13

Fields, C. (2023) Physics as Information Processing, lectures given to the Active Inference Institute, https://www.youtube.com/watch?v=RpOrRw4EhTo&t=895s

Friston K., Kilner, J. & Harrison, L. (2006) A free energy principle for the brain. J. Physiol. Paris 100, 70–87

Friston K. (2010) The free-energy principle: a unified brain theory? Nature Reviews Neuroscience

Heins R.C., Mirza M, Parr T, Friston K, Kagan I, and Pooresmailli A (2020) Deep Active Inference and Scene Construction, Frontiers in Artificial Intelligence, 3

Heinze S. et al (2023), the Insect Brain Database, https://insectbraindb.org . Curators of the database are: Berg B.G., Bucher G., el Jundi B, Farnworth M., Gruithuis J., Hartenstein V., Heinze S., Hensgen R., Homberg U., Pfeiffer K., Pfuhl G., Rossler W., Rybak J., Younger M.

Hoffman D.D. (2009) The Interface Theory of Perception (in "Object Categorization: Computer and Human Vision Perspectives," edited by Sven Dickinson, Michael Tarr, Ales Leonardis and Bernt Schiele. Cambridge University Press, 2009)

Hoffman, D.D. (2017) The Interface Theory of Perception (in The Stevens' Handbook of Experimental Psychology and Cognitive Neuroscience)

Hoffman D.D., Singh M. and Prakash, C. (2015) The Interface Theory of Perception, Psychon Bull Rev (2015) 22:1480–1506

Knill, D. C., and Pouget, A. (2004). The Bayesian brain: the role of uncertainty in neural coding and computation. *Trends Neurosci.* 27, 712–719. doi: 10.1016/j.tins.2004.10.007

Mamassian, P., Landy, M., & Maloney, L. T. (2002). Bayesian modeling of visual perception. In R. Rao, B. Olshausen, &M. Lewicki (Eds.), Probabilistic models of the brain: perception and neural function (pp. 13–36). Cambridge, MA: MIT Press

Marr, D. (1982). Vision. San Francisco: Freeman

Maynard-Smith, J. (1982). Evolution and the theory of games. Cambridge, UK: Cambridge University Press

Mirza B, Adams R, Mathys C and Friston K (2016) Scene Construction, Visual Foraging and Active Inference, Frontiers in Computational Neuroscience, 10:56

Murray, S. O., Olshausen, B. A., and Woods, D. L. (2003). Processing shape, motion and three-dimensional shape-from-motion in the human cortex. *Cereb. Cortex* 13, 508–516. doi: 10.1093/cercor/13.5.508

Parr T, Sajid N, Da Costa L, Mirza M & Friston K (2021) Generative Models for Active Vision, Frontiers in Neurobotics, 15

Prakash, C., Stephens, K. D., Hoffman, D. D., Singh, M. and Fields, C. (2021) Fitness beats truth in the evolution of perception. Acta Biotheoretica 69: 319-341 (doi: 10.1007/s10441-020-09400-0).

Prakash, C., Fields, C., Hoffman, D. D., Prentner, R., and Singh, M. (2020) Fact, fiction, and fitness. Entropy 22: 514 (doi: 10.3390/e22050514).

Rudrauf, D., Bennequin, D., Granic, I., Landini, G., Friston, K., and Williford, K. (2017). A mathematical model of embodied consciousness. J. Theor. Biol. 428, 106–131

Rudrauf D, Sergeant-Perthuis G. Belli O. and Di Marzo Serugendo G. (2022) Modeling the subjective perspective of consciousness and its role in the control of behaviours, Journal of Theoretical Biology 534

Sherman, S. M., and Guillery, R. W. (2006). Exploring the Thalamus and Its Role in Cortical Function. Cambridge, MA: MIT Press.

Smith R, Friston K, and Whyte C (2022) A step by step Tutorial on Active Inference and its Application to Empirical Data, Journal of Mathematical Psychology, 107, 102632

Solms, M. (2018) The Hard Problem of Consciousness and the Free Energy Principle, Front Psychol. 9: 2714

Tononi G. (2012) Integrated information theory of consciousness: an updated account, Archives Italiennes de Biologie, 150: 290-326

Van Der Maele T, Verbelen T, Catal O, & Dhoedt, B (2023) Embodied Object Representation Learning and Recognition, Frontiers in Neurobotics 16.

Van Der Maele T, Verbelen T, Mazzaglia P, Ferraro S, & Dhoedt, B (2023) Object-Centric Scene Representations using Active Inference, arXiv 2302.03288v1

Worden, R.P. (1999) Hybrid Cognition, Journal of Consciousness Studies, 6, No. 1, 1999, pp. 70-90

Worden, R. P. (2020a) Is there a wave excitation in the thalamus? arXiv:2006.03420

Worden, R. P. (2020b). An Aggregator model of spatial cognition. arXiv 2011.05853.

Worden R.P, Bennett M and Neascu V (2021) The Thalamus as a Blackboard for Perception and Planning, Front. Behav. Neurosci., 01 March 2021, Sec. Motivation and Reward, https://doi.org/10.3389/fnbeh.2021.633872

Worden R.P. (2024a) The Requirement for Cognition in an Equation http://arxiv.org/abs/2405.08601

Worden R.P. (2024b) Spatial Cognition: a Wave hypothesis, paper to be posted on arXiv

Worden R.P. (2024c) The Projective Wave Theory of Consciousness paper to be posted on arXiv

Worden R.P (2024d) The Evolution of Language and Human Rationality, to appear in the proceedings of the 14th EvoLang conference on the evolution of language, Madison, Wisc.; arXiv2405.06377